\documentclass[twocolumn,prb,amsmath,amssymb,floatfix,superscriptaddress]{revtex4-1}   
\usepackage{graphicx}
\usepackage{dcolumn}
\usepackage{bm}
\usepackage{xspace}
\usepackage{hyperref}
\usepackage{color}
\usepackage[flushleft]{threeparttable}
\begin{document}

\title{Evidence for a Dirac nodal-line semimetal in SrAs$_{3}$}

\author{Shichao~Li}
\author{Zhaopeng~Guo}
\author{Dongzhi~Fu}
\author{Xing-Chen~Pan}
\author{Jinghui~Wang}
\author{Kejing~Ran}
\author{Song~Bao}
\author{Zhen~Ma}
\author{Zhengwei~Cai}
\affiliation{National Laboratory of Solid State Microstructures and Department of Physics, Nanjing University, Nanjing 210093, China}
\author{Rui~Wang}
\affiliation{Department of Physics and Astronomy, Shanghai Jiao Tong University, Shanghai 200240, China}
\author{Rui Yu}
\affiliation{School of Physics and Technology, Wuhan University, Wuhan 430072, China}
\author{Jian Sun}
\author{Fengqi Song}
\email{songfengqi@nju.edu.cn}
\author{Jinsheng~Wen}
\email{jwen@nju.edu.cn}
\affiliation{National Laboratory of Solid State Microstructures and Department of Physics, Nanjing University, Nanjing 210093, China}
\affiliation{Collaborative Innovation Center of Advanced Microstructures, Nanjing University, Nanjing 210093, China}

\maketitle

\noindent {\bf Abstract}\\
Dirac nodal-line semimetals with the linear bands crossing along a line or loop, represent a new topological state of matter. Here, by carrying out magnetotransport measurements and performing first-principle calculations, we demonstrate that such a state has been realized in high-quality single crystals of SrAs$_{3}$. We obtain the nontrivial $\pi$ Berry phase by analysing the Shubnikov-de Haas quantum oscillations. We also observe a robust negative longitudinal magnetoresistance induced by the chiral anomaly. Accompanying first-principles calculations identify that a single hole pocket enclosing the loop nodes is responsible for these observations.


\noindent {\bf 1. Introduction}\\
Recently, a new type of topological materials, including three-dimensional (3D) Dirac, Weyl, and nodal-line semimetals, have attracted huge interests\cite{wehling2014dirac,nm15_1145}.
The Dirac semimetals are analogues to graphene in 3D, with the inverted linear bands crossing at the Dirac nodes near the Fermi level\cite{PhysRevB.85.195320}. By breaking either the spatial-inversion ($P$) or time-reversal ($T$) symmetry both present in Dirac semimetals, a four-fold degenerated Dirac node
splits into two Weyl nodes with opposite chirality, and these materials are termed Weyl semimetals\cite{PhysRevB.83.205101,huang2015weyl,PhysRevX.5.011029}.
The low-energy physics of Dirac and Weyl semimetals are described by a Dirac and Weyl equation, respectively\cite{wehling2014dirac}. Both Dirac\cite{Liu21022014,nc5_3786,liang2015ultrahigh} and Weyl semimetals\cite{huang2015weyl,xu2015discovery,lv2015observation,scpmaguo} have been established, and shown to exhibit a number of intriguing transport properties, such as the large longitudinal magnetoresistance (MR) and high mobility\cite{ali2014large,liang2015ultrahigh,shekhar2015extremely,scpmadu}, nontrivial $\pi$ Berry phase\cite{PhysRevB.96.041201,PhysRevLett.113.246402,huang2015observation,csbwang}, and chiral-anomaly-induced negative longitudinal MR~\cite{huang2015observation,xiong2015evidence,np12_550,nc7_10735,PhysRevLett.118.096603,PhysRevB.88.104412,PhysRevB.93.121112}.
Topological nodal-line semimetals, where the linear bands cross each other along a line or loop, instead of at discrete points as in Dirac and Weyl semimetals\cite{PhysRevB.84.235126,PhysRevB.92.045108,cpb117106,csbakira}, have also been proposed in various systems, including ZrSi$X$ ($X$=S, Se, Te)~(refs~\onlinecite{schoop2016dirac,PhysRevLett.117.016602,singha2017large}), Cu$_3X$N ($X$=Pd, Zn)~(ref.~\onlinecite{PhysRevLett.115.036807}), $X$TaSe$_2$ ($X$=Pb, Tl)~(ref.~\onlinecite{bian2016topological}), PtSn$_4$~(ref.~\onlinecite{wu2016dirac}), Ca$TX$ ($T$=Cd, Ag; $X$=P, Ge, As)~(ref.~\onlinecite{PhysRevB.95.245113}), Ca$_3$P$_2$~(ref.~\onlinecite{PhysRevB.93.205132}), and CaP$_3$ family~(refs~\onlinecite{PhysRevB.95.045136,PhysRevLett.118.176402}). However, many of these proposals, such as the topological nodal-line state in the CaP$_3$ family, are calling for experimental verification.

\begin{figure*}[htb]
\centering
\includegraphics[width=\linewidth]{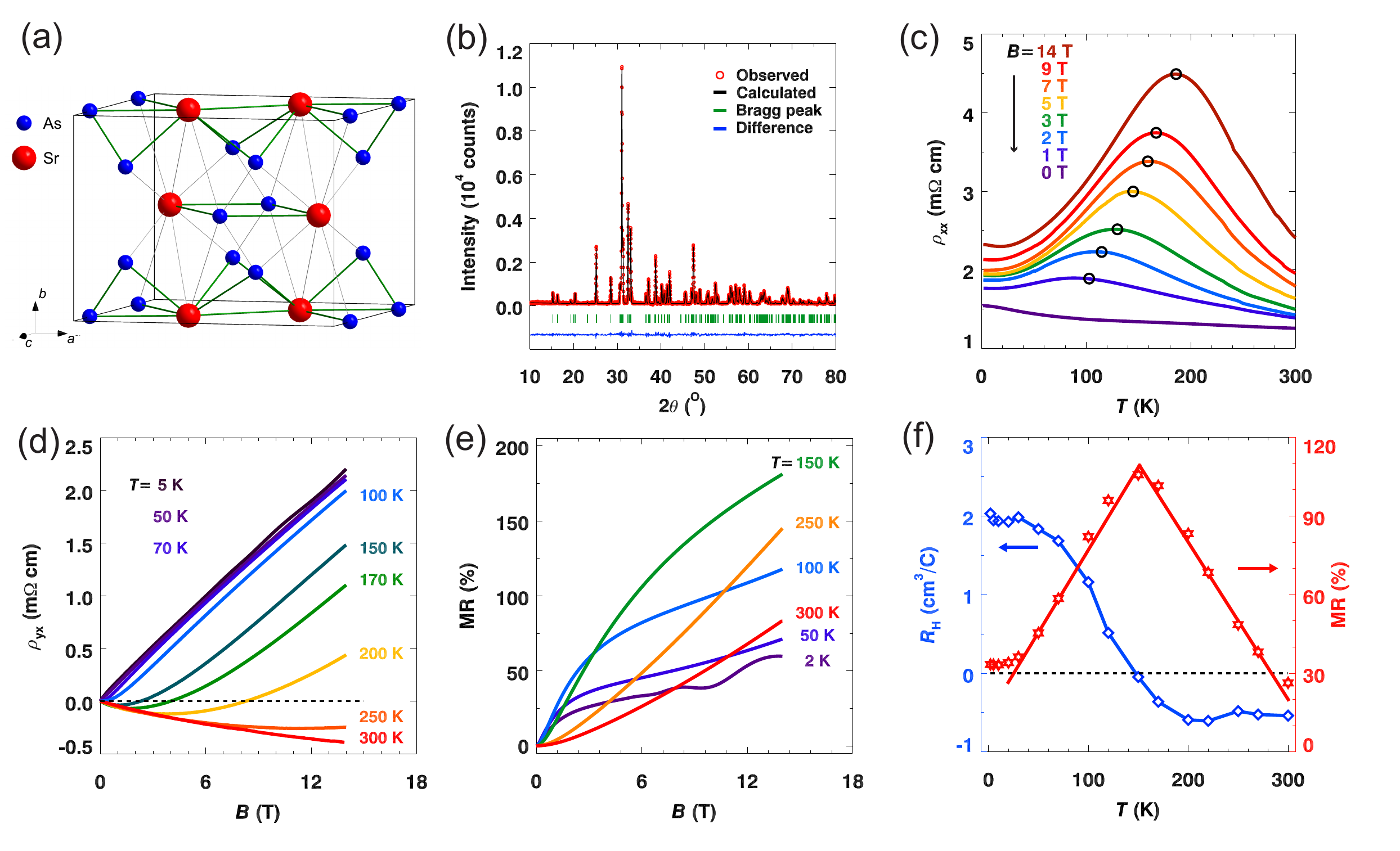}
\caption{(Color online) 
{\bf Crystal structure and longitudinal magnetoresistance.}
{{\bf a} Schematic for the crystal structure of SrAs$_{3}$ (monoclinic, space group C2/m, No. 12). {\bf b} X-ray diffraction patterns measured at room temperature and the refinement results. {\bf c} Temperature dependence of the resistivity ($\rho_{xx}$) under different magnetic fields. Circles indicate the positions of the turning point at $T^{*}$. {\bf d} and {\bf e} Magnetic-field dependence of the Hall resistivity $\rho_{yx}$ and magnetoresistance (MR), respectively. {\bf f} Left, Hall coefficient $R_{H}(T)$ extracted at $B=$1~T; Right, MR$(T)$ extracted at $B=$6~T.}
\label{fig:electrical}
}
\end{figure*}

The crystal structure of the CaP$_3$ family contains puckered polyanionic layers stacking along $b$ axis, as illustrated in Fig.~\ref{fig:electrical}a~(ref.~\onlinecite{bauhofer1981structure}). The material of interest in this work, SrAs$_{3}$, crystallizes into the monoclinic structure with the C2/m space group\cite{bauhofer1981structure}. SrAs$_{3}$ was previously known as a narrow-gap semimetal\cite{bauhofer1981structure}, but a very recent theory work suggested that it was a Dirac nodal-line semimetal protected by $PT$ and mirror symmetries, if the spin-orbit coupling (SOC) effect was neglected\cite{PhysRevB.95.045136}. Besides, several other members in the CaP$_3$ family, such as CaP$_3$, CaAs$_3$, SrP$_3$, and BaAs$_3$ were also predicted to be such topological semimetals\cite{PhysRevB.95.045136,PhysRevLett.118.176402}. It is highly desirable to realize the predicted topological state in these materials experimentally.

Here, by measuring the magnetoresistance on high-quality single crystals of SrAs$_3$, we observe the nontrivial $\pi$ Berry phase by analysing the Shubnikov-de Haas (SdH) quantum oscillation data, and the robust negative MR induced by the chiral anomaly. First-principles calculations show that a single hole pocket enclosing the loop nodes is responsible for these exotic properties. These results unequivocally demonstrate that SrAs$_3$ is a Dirac nodal-line semimetal as proposed in ref.~\onlinecite{PhysRevB.95.045136}.

\begin{figure}[htb]
\centering
\includegraphics[width=\linewidth,trim=0mm 0mm 0mm 0mm,clip]{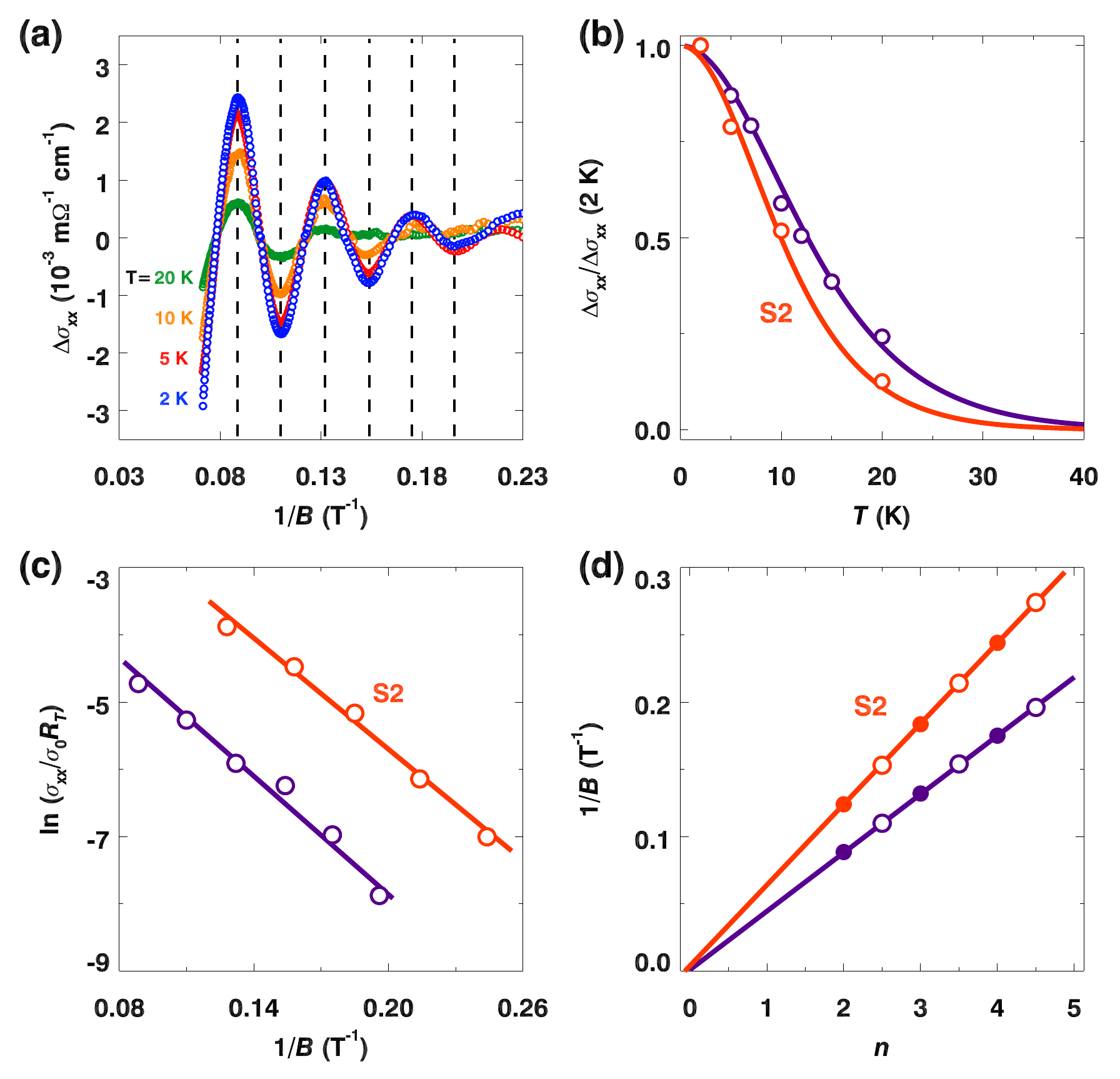}
\caption{(Color online) 
{\bf SdH oscillation and nontrivial Berry phase.}
{{\bf a} Oscillatory components $\Delta\sigma_{xx}$ at four temperatures. Dashed lines indicate the peaks and valleys. {\bf b} Temperature dependence of the oscillatory amplitude, normalized by $\Delta\sigma_{xx}$ at 2~K. Lines through data are fits described in the main text. {\bf c} Fits of $R_{\rm T}$ to extract Dingle temperatures. {\bf d} Landau-level fan diagram. Closed and open circles denote integers ($\Delta\sigma_{xx}$ peak) and half integers ($\Delta\sigma_{xx}$ valley), respectively. Lines through data are linear fits. In {\bf b}, {\bf c}, and {\bf d}, the results for an additional sample labelled as S2 are also shown. }
\label{fig:topological}
}
\end{figure}

\begin{figure*}[htb]
\centering
\includegraphics[width=1\linewidth,trim=0mm 0mm 0mm 0mm,clip]{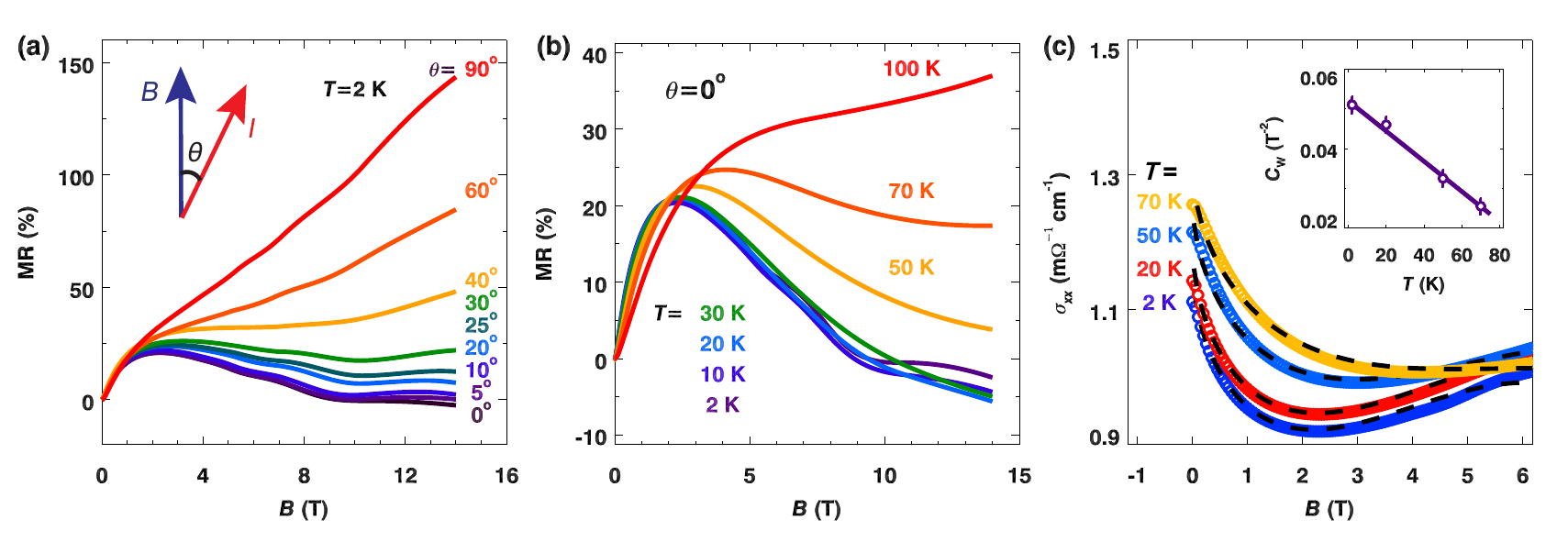}
\caption{(Color online) 
{\bf Negative longitudinal MR and chiral anomaly.}
{{\bf a} Longitudinal MR measured at different angles ($\theta$) between magnetic field ($B$) and electrical current ($I$) at $T=2$~K, and {\bf b} at different temperatures at $\theta=0^{\circ}$}. {\bf c} Longitudinal Magnetoconductivity measured at different temperatures. Dashed lines are fits to data described in the main text. The inset shows the temperature dependence of the parameter characterizing the chiral effect. The line in the inset is a guide to the eye. Errors represent one standard deviation.
\label{fig:nmr}
}
\end{figure*}

\begin{figure}[htb]
\centering
\includegraphics[width=1\linewidth]{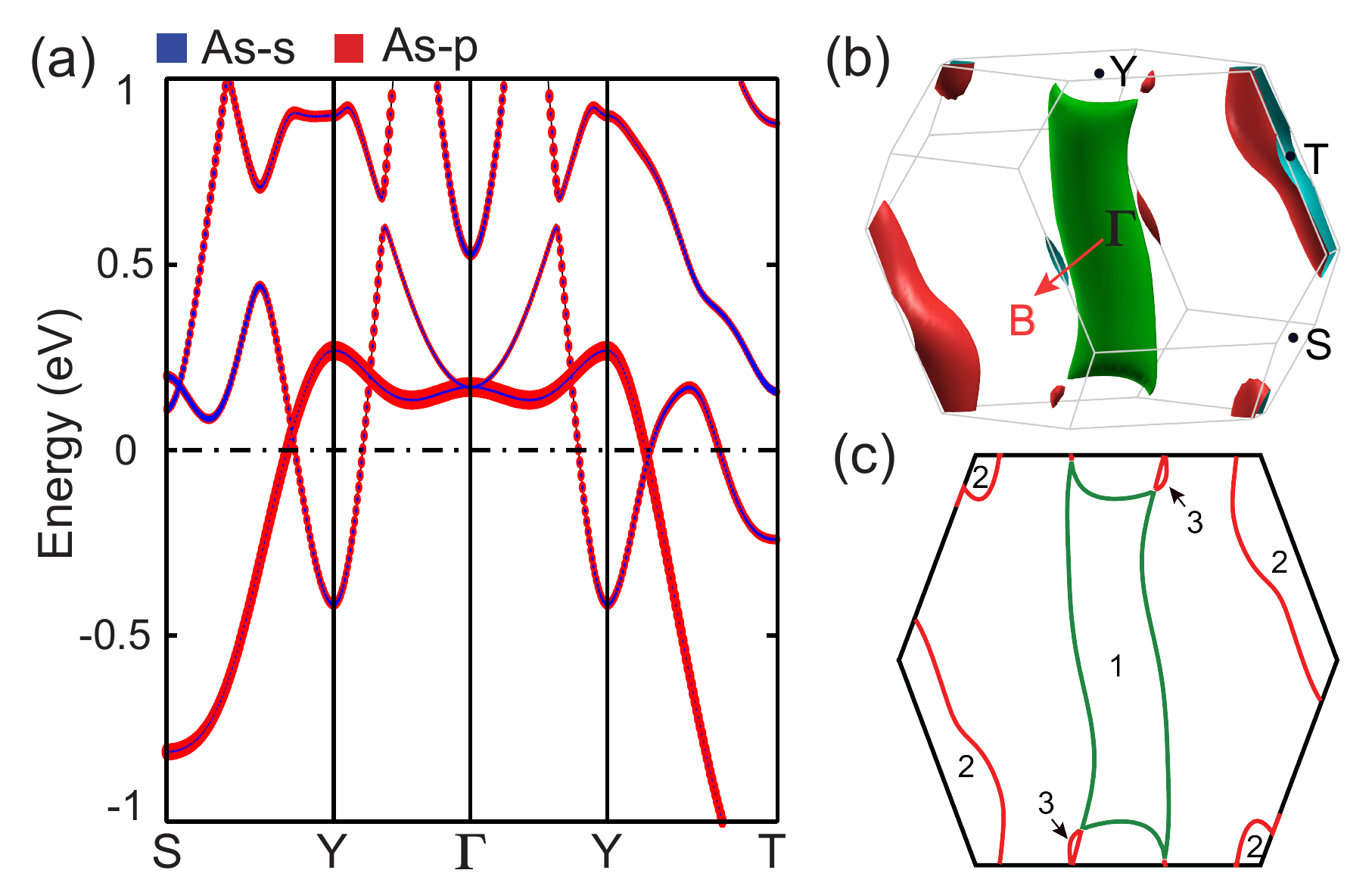}
\caption{(Color online) 
{\bf Band structure and Fermi surface.}
{{\bf a} Calculated band structure of SrAs$_{3}$. {\bf b} Calculated Fermi surface. The arrow denotes the magnetic-field direction  perpendicular to the $a$-$c$ plane. {\bf c} Cut of b in the $a$-$c$ plane. Fermi pockets 1 represent electron pockets, and 2 and 3 represent hole pockets.}
\label{fig:fs}
}
\end{figure}

\begin{figure*}[htb]
\centering
\includegraphics[width=1\linewidth]{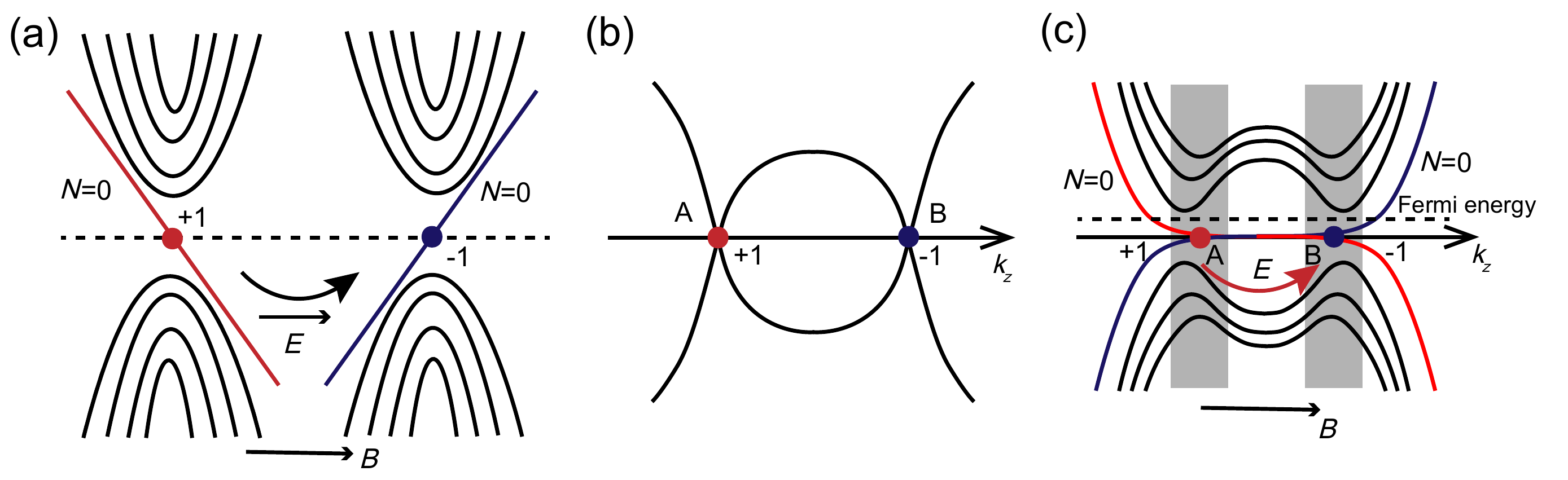}
\caption{(Color online) 
{\bf Landau Levels in Weyl and nodal-line semimetals.}
{{\bf a} Landau levels of Weyl semimetals in the presence of parallel electric ($E$) and magnetic fields ($B$). $\pm$1 stand for the opposite chirality of the N=0 Landau levels. {\bf b} Schematic of the band structure along $k_z$ in nodal-line semimetals. {\bf c} Landau levels in nodal-line semimetals after Landau quantization. In the gray shaded region, two chiral Landau level bands occur.}
\label{fig:chiral}
}
\end{figure*}

\noindent {\bf 2. Methods}\\
\noindent{\bf 2.1 Sample growth and electrical transport.} Single crystals of SrAs$_{3}$ were grown by melting stoichiometric amounts of Sr and As at 850~$^\circ$C. After 24 hours, the melt was cooled to 750~$^\circ$C at a rate of 4~$^\circ$C/h. Then the furnace was shut down and cooled to room temperature. Shiny crystals with size up to 5$\times$3$\times$1~mm$^3$ could be obtained so. The crystals have $a$-$c$ plane as the cleavage plane, determined using a Laue x-ray diffractometer, consistent with the stacking arrangement in this material. The composition and structure of the single crystals were measured using a scanning electron microscope equipped with an energy-dispersive x-ray spectrometer (EDS), and a powder x-ray diffractometer, respectively. Refinements of the x-ray diffraction (XRD) data were performed with the FullProf. program. Magnetotransport measurements were performed in a Physical Property Measurement System with a standard six-contact method, which enabled the measuring of the electrical and Hall resistivity simultaneously. The contacts were made on the $a$-$c$ plane of the bar-shape sample.

\noindent{\bf 2.2 DFT calculations.} First-principles calculations were performed by using the projected augmented wave method implemented in the Vienna ab-initio simulation package\cite{PhysRevB.54.11169} based on the generalized gradient approximation in the Perdew-Burke-Ernzerhof functional theory\cite{PhysRevLett.77.3865}. The energy cutoff of 310~eV was set for the plane-wave basis and a $k$-point mesh of 7$\times$7$\times$7 was used.

\noindent {\bf 3. Results}\\

\noindent{\bf 3.1 Sample characterization.} We first check the composition of the SrAs$_3$ single crystals and confirm that the crystals are stoichiometric with the molar ratio of Sr:As=1:3. In Fig.~\ref{fig:electrical}b, we show the XRD data for SrAs$_3$ powders obtained by grinding the single crystals. It clearly demonstrates that the sample contains a single phase. The XRD pattern can be well indexed with the monoclinic structure (space group C2/m, No.~12), as illustrated in Fig.~\ref{fig:electrical}a. From the refinements, we obtain the lattice constants $a=9.60(8)$~\AA, $b=7.65(8)$~\AA, and $c= 5.86(9)$~\AA, and $\alpha=\gamma=90^\circ$ and $\beta=112.87(0)^\circ$. These results are consistent with the existing literature\cite{bauhofer1981structure}.

\noindent{\bf 3.2 Longitudinal magnetoresistance and Hall resistivity.} In Fig.~\ref{fig:electrical}c, we plot the temperature dependence of the resistivity ($\rho_{xx}$) under different magnetic fields for SrAs$_{3}$ with field $B$ applied perpendicular to the electrical current $I$. The current flows in the $a$-$c$ plane, not along any particular axis. Under zero field, the temperature dependence of $\rho_{xx}$ is semiconductor-like, but the value of the resistivity is low. These results agree with previous reports on this material\cite{bauhofer1981structure,PhysRevB.30.1099}. Interestingly, when we apply a field, the material undergoes a transition from semiconductor at high temperatures to metal at low temperatures, at a temperature we label as $T^*$. As can be seen from Fig.~\ref{fig:electrical}c, $T^*$ increases monotonically with the field. We believe that this transition is due to the change of the carrier type. In Fig.~\ref{fig:electrical}d and f, a sign change of the Hall resistivity $\rho_{yx}$ and Hall coefficient $R_{\rm H}$ from positive (hole) to negative (electron) at $T^*$ upon heating can be clearly observed. Similar behaviors were also suggested in ref.~\onlinecite{PhysRevB.30.1099,zhou1984determination,PhysRevB.50.5180}. These observations can be understood by looking at the band structure in Fig.~\ref{fig:fs}a, where there are some very flat bands around the $\Gamma$ point that are very close to the Fermi level. 
These flat bands have a large density of states but the electron mobility in these bands is low. In zero field, these bands dominate the transport properties over the metallic hole bands that have low density of states, which gives rise to the semiconducting behavior as shown in Fig.~\ref{fig:electrical}c. In the magnetic fields, at high temperatures, the electrons in these bands can be activated to the Fermi level, but their mobility is rather low, and thus the non-metallic transport behavior is observed. At low temperatures, the Dirac-type hole band dominates the transport, and gives rise to the metallic properties, consistent with the positive $R_H$ at low temperatures shown in Fig.~\ref{fig:electrical}f.
From the results in Fig.~\ref{fig:electrical}c and d, the hole concentration and mobility at 2~K are estimated to be $3.07(9)\times10^{18}$~cm$^{-3}$ and $1.35(1)\times10^3$~cm$^2$/Vs, respectively. These values are listed in Table~\ref{tab:para}, together with some other important quantities for this material. The change of the carrier type also leads to the different temperature dependence of the magnetoresistance~[MR=$(\rho_{xx}(B)-\rho_{xx}(0))/\rho_{xx}(0)$], as we plot in Fig.~\ref{fig:electrical}e and f. Throughout the paper, the MR data have been symmetrized to the magnetic field to get rid of the contribution from the Hall resistivity. At low temperatures, MR increases with temperature and then decreases above $T^*$, as clearly shown in Fig.~\ref{fig:electrical}f. Note that, as shown in Fig.~\ref{fig:electrical}e, the Hall resistivity shows linear, positive field dependence below 100 K, which indicates that the transport properties are dominated by holes. Therefore, we use the one-band model to discuss the transport properties at low temperatures throughout the paper.

\noindent{\bf 3.3 Quantum oscillation data.} In Fig.~\ref{fig:electrical}e, we have already observed oscillating behaviors of the MR at 2~K. To further proceed, we convert the resistivity $\rho_{xx}$ into conductivity $\sigma_{xx}$, using $\sigma_{xx}=\rho_{xx}/(\rho_{xx}^2+\rho_{yx}^2)$, which is a more appropriate way in analysing the SdH data when the condition of $\rho_{xx}\gg\left|\rho_{yx}\right|$ is not satisfied\cite{doi:10.7566/JPSJ.82.102001}. The SdH oscillatory components $\Delta\sigma_{xx}$ at several temperatures, obtained by subtracting the smooth background, are plotted as a function of $1/B$ in Fig.~\ref{fig:topological}a. The oscillation is obvious even at a modest field of 6~T at low temperatures. We observe a single oscillation frequency, indicating only one Fermi pocket participates in the oscillation process. The oscillatory components are often described using the Lifshitz-Kosevich formula\cite{liang2015ultrahigh,murakawa2013detection,doi:10.7566/JPSJ.82.102001,PhysRevLett.113.246402,zmhcpb}, \begin{eqnarray}\label{SdH1}
\Delta\sigma_{xx}\propto R_{\rm T}R_{\rm D}R_{\rm S}{\rm cos}[2\pi(F/B-\gamma+\delta)],
\end{eqnarray}
where Onsager phase factor $\gamma=1/2-\phi_{\rm B}/2\pi$, and $\delta$ is a phase shift. $F$ and $\phi_{\rm B}$ are the oscillation frequency and Berry phase, respectively. $R_{\rm T}=(2\pi^{2}k_{\rm B}T/\hbar\omega_{c})/\sinh(2\pi^{2}k_{\rm B}T/\hbar\omega_{c})$, and $R_{\rm D}=\exp(-2\pi^{2}k_{\rm B}T_{\rm D}/\hbar\omega_{c})$, with $\omega_{c}$ and $T_{\rm D}$ being the  cyclotron frequency and Dingle temperature, respectively. $R_{\rm S}=\cos(\frac{1}{2}\pi gm^*/m_0)$ is spin damping factors due to the Zeeman splitting, which can be neglected in the present case, as we do not observe any sign of Landau level splitting in the SdH oscillations at $T=2$~K under $B=$14~T. Here, $m^*$ and $m_0$ are the effective mass and rest mass of the free electron. By analysing Fig.~\ref{fig:topological}a with Eq.~\ref{SdH1}, we can extract a number of important parameters for the material, which are listed in Table~\ref{tab:para}.

Since the cyclotron frequency $\omega_{c}=eB/m^{*}$, we have $2\pi^{2}k_{\rm B}T/\hbar\omega_{c}\approx14.69m^*T/B$, and thus $R_{\rm T}$ is the only temperature-dependent term in $\Delta\sigma_{xx}$~(ref.~\onlinecite{liang2015ultrahigh,murakawa2013detection,doi:10.7566/JPSJ.82.102001}). We have fitted the temperature dependence of the amplitude of $\Delta\sigma_{xx}$, and the results are shown in Fig.~\ref{fig:topological}b. We obtain an effective mass of 0.13(3)$m_0$ for the sample by averaging the results obtained at several fields. In a second sample (S2), we observe a similar $m^*=0.12(4)m_0$. These values are  comparable to those of NbP\cite{PhysRevB.93.121112} and ZrSiS\cite{singha2017large}, indicating that the dispersion is Dirac-like with light effective mass. Dingle temperature $T_{\rm D}$ can be extracted by fitting the amplitude of the SdH oscillations at a fixed temperature under different magnetic fields. The results at $T=2$~K are presented in Fig.~\ref{fig:topological}c, which yield $T_{\rm D}=$7.70(5)~K. Through $\tau_q=\hbar/(2\pi k_{\rm B}T_{\rm D})$ and $\mu_q=e\tau/m^*$~(ref.~\onlinecite{PhysRevB.95.245113}), we obtain the quantum scattering time $\tau_q=0.16$~ps and quantum mobility $\mu_{\rm q}=2080$~cm$^2$/Vs. By performing similar analysis, we obtain $T_{\rm D}=7.40(3)$~K for S2.

According to the Onsager quantization rule\cite{murakawa2013detection,doi:10.7566/JPSJ.82.102001}, by assigning a maximum of $\Delta\sigma_{xx}$ to an integer $n$, we can construct the Landau-level fan diagram, as shown in Fig.~\ref{fig:topological}d. In this case, $2\pi(F/B+\gamma-\delta)=n\times2\pi$, and the slope and intercept of $1/B$ $vs$ $n$ correspond to $1/F$ and $\gamma-\delta$, respectively. From Fig.~\ref{fig:topological}d, we extract a single oscillation frequency of $F=23.25(6)$~T at 2~K for the main sample, corresponding to a pocket area of $S=2.21(5)\times10^{-3}$~\AA$^{-2}$. The Fermi energy and velocity associated with the bands are 41.76(1)~meV and $2.36(2)\times10^{5}$~m/s, respectively. The intercept $\phi_{\rm B}/2\pi-1/2=0.02$ gives $\phi_{\rm B}=1.04(4)\pi$ and $\delta=0$, which is similar to that in refs~\onlinecite{PhysRevLett.117.016602,singha2017large,huang2015observation}.
Similar practices for a second sample result in $\phi_{\rm B}=1.09(6)\pi$. These results clearly show that the band structure is topologically nontrivial.

\begin{table}[htb]
\centering
\caption{Parameters of the SrAs$_3$ single crystal obtained at 2~K. $n_h$, hole concentration; $\mu_h$, mobility; $m^*/m_0$, effective mass over the rest mass of the free electron; $F$, oscillation frequency; $S$, area of the Fermi pocket; $E_{\rm F}$, Fermi energy; $v_{\rm F}$, Fermi velocity; $\phi_{\rm B}$, Berry phase; $T_{\rm D}$, Dingle temperature.}\label{tab:para}
\begin{tabular*}{0.5\textwidth}{@{\extracolsep{\fill}}cc}\hline \hline
\begin{minipage}{3cm}\vspace{1mm} \centering $n_h$ (cm$^{-3}$) \vspace{1mm} \end{minipage} & $3.07(9)\times10^{18}$ \\
\begin{minipage}{3cm}\vspace{1mm} \centering $\mu_h$ (cm$^{2}$/Vs) \vspace{1mm} \end{minipage} & $1.35(1)\times10^3$  \\
\begin{minipage}{3cm}\vspace{1mm} \centering $m^*/m_0$ \vspace{1mm} \end{minipage} & 0.13(3)\\
\begin{minipage}{3cm}\vspace{1mm} \centering $F$ (T) \vspace{1mm} \end{minipage} & 23.25(6)\\
\begin{minipage}{3cm}\vspace{1mm} \centering $S$ (\AA$^{-2}$) \vspace{1mm} \end{minipage} & $2.21(5)\times10^{-3}$ \\
\begin{minipage}{3cm}\vspace{1mm} \centering $E_{\rm F}$ (meV) \vspace{1mm} \end{minipage} & 41.76(1) \\
\begin{minipage}{3cm}\vspace{1mm} \centering $v_{\rm F}$ (m/s) \vspace{1mm} \end{minipage} & $2.36(2)\times10^5$ \\
\begin{minipage}{3cm}\vspace{1mm} \centering $\phi_{\rm B}$ \vspace{1mm} \end{minipage} & $1.04(4)\pi$
\\
\begin{minipage}{3cm}\vspace{1mm} \centering $T_{\rm D}$ (K) \vspace{1mm} \end{minipage} & 7.70(5)\\
\hline \hline\end{tabular*}
\end{table}

\noindent{\bf 3.4 Negative longitudinal MR and chiral anomaly.} Another strong evidence for the topological nodal structure is the observation of chiral-anomaly-induced negative MR\cite{wehling2014dirac,xiong2015evidence,np12_550}. In topological semimetals, a chiral current can be induced between two Weyl points with opposite chirality for $B||I$. In Fig~\ref{fig:nmr}a and b, we show the magnetic-field dependence of MR at different angles and temperatures, respectively. Negative MR can be observed for $\theta\le30^\circ$ at 2~K~(Fig~\ref{fig:nmr}a), and at $T\le70$~K for $\theta=0^\circ$~(Fig~\ref{fig:nmr}b). The negative MR can be analysed using the semiclassical formula\cite{huang2015observation,singha2017large,PhysRevLett.118.096603},
\begin{eqnarray}\label{nmr}
\sigma(B)=(1+CB^2)\sigma_{\text{WAL}}+\sigma_{\text{N}},
\end{eqnarray}
where $C$ is a temperature-dependent positive parameter originating from the chiral anomaly, and $\sigma_{\text{WAL}}=\sigma_{0}+a\sqrt{B}$ and $\sigma_{\text{N}}=(\rho_{0}+AB^{2})^{-1}$ are the conductivity resulting from the weak-antilocalization effect and conventional non-linear band contributions, respectively. The experimental data can be well fitted with Eq.~\ref{nmr}, as shown in Fig.~\ref{fig:nmr}c. The chiral-anomaly effect decreases with increasing temperature, as demonstrated by the monotonic decrease of $C$ in temperature. Our observations of the negative MR in SrAs$_3$ strongly suggest that it is a topological nodal-line semimetal, where chiral anomaly plays a role in the transport when $B||I$. We have also measured the magnetoresistance with current flowing in a direction perpendicular to the present setup and obtained similar results, demonstrating that our results do not depend on the crystalline axis.

\noindent{\bf 3.5 Band structure.} Our first-principles calculation results are shown in Fig.~\ref{fig:fs}. In Fig.\ref{fig:fs}a, we clearly observe that the linear inverted bands cross each other around the Y point near the Fermi level, and these crossing points form a nodal loop, consistent with previous results in ref.~\onlinecite{PhysRevB.95.045136}. Since the nodal loop is tilted with respect to the Fermi level, we expect a large $\theta$ range where negative MR can be observed. Compared to the results in ref.~\onlinecite{PhysRevB.95.045136}, the nearly-flat bands around the $\Gamma$ point are much closer to the Fermi level. This is fully consistent with our observations on the carrier-type change upon changing the temperature.

We plot the calculated Fermi surface in Fig.\ref{fig:fs}b, and a cut of it in the $a$-$c$ plane in Fig.~\ref{fig:fs}c. The cut consists of three sets of Fermi pockets marked by 1, 2 and 3, with 1 being the electron pocket, and 2 and 3 being the hole pockets.
The area of pocket 3 is calculated to be 2.05(4)$\times$10$^{-3}$~\AA$^{-2}$, very close to experimental value of $2.21(5)\times10^{-3}$~\AA$^{-2}$. Thus, the calculation results firmly prove the presence of a small hole pocket enclosing the nodal points. Such a topological nodal structure is fully compatible with our magnetotransport results.

\noindent {\bf 4. Discussion}\\
The observation of the negative MR in the nodal-line semimetal SrAs$_3$ could possibly be a result of chiral anomaly similar to that in Weyl semimetals. In Weyl semimetals, the chiral anomaly can be understood from the Landau bands after quantization. As shown in the schematic plot in Fig.~\ref{fig:chiral}a, two $N = 0$ Landau bands with opposite chirality occur. In this case, a parallel electric field induces the shift of electrons from one Weyl valley to the other, resulting in the negative MR. For nodal-line semimetals, we use a low-energy effective model, 
\begin{eqnarray}\label{hamitonian}
H_{\rm eff}=v\tau_{x}\bm{\sigma}\cdot {\bf p}+b \tau_{z}\sigma_{x} ,
\end{eqnarray}
where, {$\bm{\sigma}$} and $\bm{\tau}$ are the Pauli matrix denoting spin and orbital degrees of freedom respectively, to illustrate the possible negative MR. This Hamiltonian enjoys a gapless nodal loop in $k_y$-$k_z$ plane, {\it i.e.}, $\sqrt{k_{y}^2+k_{z}^2}=b/v$. Fig.~\ref{fig:chiral}b shows the spectrum along $k_z$ with $k_x= k_y= 0$ and two crossing points $A = (0,\, 0,\,-v/b)$ and $B=(0,\,0,\,v/b)$. Then, we consider a magnetic field along $z$ and study the Landau quantization. Interestingly, it is found that two $N = 0$ Landau bands emerge which become degenerate within the region $k_z\in [-v/b,\,v/b]$, as shown by the red and dark blue curves in Fig.~\ref{fig:chiral}c. Furthermore, outside this region, the two bands are both chiral and have opposite chirality, {\it i.e.}, with opposite Fermi velocity for a certain $k_z$. This result is similar to that in a Weyl semimetal case (Fig.~\ref{fig:chiral} a), where chiral anomaly will play an important role when $E||B$, giving rise to the charge pumping effect and negative MR in nodal-line semimetals.

Besides chiral anomaly, there are other origins that can also give rise to negative MR. First, current jetting effect can induce negative MR when inhomogeneous currents are injected into materials with high mobility. The low mobility in SrAs$_3$ makes this scenario unlikely. In addition, the electrical contacts were made covering the entire sample width to avoid this. Second, weak localization can cause negative MR at low fields with $\sim B^{-1/2}$ dependence\cite{xiong2015evidence}, inconsistent with our observations. Moreover, the negative MR resulting from this effect should be hardly dependent on the angle between $B$ and $I$. Third, conductivity fluctuation mechanism\cite{PhysRevB.95.241113} can also lead to negative MR, which exists at very high temperatures, while we only observe the negative MR up to 70~K. Finally, our conclusion that the negative MR in SrAs$_3$ is induced by the chiral anomaly is consistent with the nontrivial $\pi$ Berry phase. Our own calculations and those in ref.~\onlinecite{PhysRevB.95.045136} further support that this material is a topological nodal-line semimetal.

In previous calculations\cite{PhysRevB.95.045136,PhysRevLett.118.176402}, the Dirac nodal structure in the CaP$_3$ family is present only when the SOC effect is neglected. When SOC is taken into account, a small gap will open and the system evolves into a topological insulator\cite{PhysRevB.95.045136,PhysRevLett.118.176402}. Our results show that the linear bands contribute to the electrical transport in the real material. We speculate that either there are some additional symmetries protecting the Dirac points, such as the crystal symmetry\cite{wehling2014dirac}, thus preventing the opening of a gap, or a small gap indeed opens, but the linear dispersion still persists\cite{pnas114_816}. We anticipate our work to stimulate angle-resolved photoemission spectroscopy and scanning tunneling microscopy studies to provide more insights. Under an external magnetic field, the time-reversal symmetry that protects a Dirac point is broken. As a result, a Dirac point should split into two Weyl points with opposite chirality\cite{wehling2014dirac,PhysRevB.83.205101}. The distance between the two Weyl points in the momentum space depends on the strength of the field. In our work, we observe only one oscillation frequency, signifying only a single set of Fermi pocket enclosing the loop nodes up to 14~T. It will be interesting to study this material under higher fields to check whether additional pockets will emerge.



\bigskip
\noindent {\bf Acknowledgements}\\
We thank Hongming Weng, Yiming Pan, Shao-Chun Li, and Wentao Zhang for stimulating discussions. The work was supported by NSFC (Grant Nos 11674157, 51372112 and 11574133), and by the MOST of China (Grant Nos 2016YFA0300404 and 2015CB921202).

\noindent {\bf Author contributions}\\
J.S.W. and F.Q.S. conceived the project. S.C.L. grew the single crystals, S.C.L. and D.Z.F. performed the magnetotranport measurements. Z.P.G, J.S. and R.Y. carried out the first principle calculations. S.C.L., X.C.P. and J.S.W. analysed the data. J.S.W. and S.C.L. wrote the paper with comments from all authors.

\medskip
\noindent {\bf Correspondence}\\
\noindent Correspondence and request for materials should be addressed to F.Q.S.~(Email: \mbox{songfengqi@nju.edu.cn}), or J.S.W.~(Email: \mbox{jwen@nju.edu.cn}).

\end{document}